# Selective formation of metastable polymorphs in solid-state synthesis

Yan Zeng[1,†], Nathan J. Szymanski[1,2,†], Tanjin He[1,2], KyuJung Jun[1,2], Leighanne C. Gallington[3], Haoyan Huo[1,2], Christopher J. Bartel[4], Bin Ouyang[5], Gerbrand Ceder[1,2,*]

**Affiliations:**

[1]Materials Sciences Division, Lawrence Berkeley National Laboratory, Berkeley, CA 94720, USA
[2]Department of Materials Science and Engineering, University of California Berkeley, Berkeley, CA 94720, USA
[3]X-ray Science Division, Argonne National Laboratory, Lemont, Illinois 60439, USA
[4]Department of Chemical Engineering and Materials Science, University of Minnesota, Minneapolis, MN 55455, USA
[5]Department of Chemistry and Biochemistry, Florida State University, Tallahassee, FL 32306, USA
*Corresponding author: Email: gceder@berkeley.edu
†These authors contributed equally to this work

## Abstract

Metastable polymorphs often result from the interplay between thermodynamics and kinetics. Despite advances in predictive synthesis for solution-based techniques, there remains a lack of methods to design solid-state reactions targeting metastable materials. Here, we introduce a theoretical framework to predict and control polymorph selectivity in solid-state reactions. This framework presents reaction energy as a rarely used handle for polymorph selection, which influences the role of surface energy in promoting the nucleation of metastable phases. Through *in situ* characterization and density functional theory calculations on two distinct synthesis pathways targeting LiTiOPO$_4$, we demonstrate how precursor selection and its effect on reaction energy can effectively be used to control which polymorph is obtained from solid-state synthesis. A general approach is outlined to quantify the conditions under which metastable polymorphs are experimentally accessible. With comparison to historical data, this approach suggests that using appropriate precursors could enable the synthesis of many novel materials through selective polymorph nucleation.

**Introduction**

Creating specific polymorphs through targeted synthesis remains one of the great unsolved challenges of rational materials design. Metastable compounds often exhibit desirable properties for advanced technologies related to pharmaceuticals, semiconductors, catalysis, and energy. However, their synthesis is challenging due to competition with lower-energy phases[1,2], and there are few guidelines available to understand which metastable compounds are experimentally accessible. Thermodynamic competition for metastable phases may appear as decomposition into phases with different compositions (phase separation) or as a phase at the same composition but with a different structure that exhibits lower free energy (polymorphism)[3]. A compound that is metastable with respect to phase separation can sometimes be retained by keeping the synthesis temperature low, thereby restricting the long-range diffusion needed to form the competing phases[4] – an approach that has been used to synthesize metastable compounds in thin films[5]. In contrast, polymorphism remains difficult to control. The ground state for a given composition often competes with many different structures in a narrow energy range[2], and it is not well understood what determines the specific conditions under which each structure can form.

In this work, we demonstrate that the selectivity of competing polymorphs formed during solid-state synthesis can be quantitatively controlled through rational selection of precursors and their associated reaction energy to form a given phase. We show that a larger reaction energy can amplify the relative influence of surface energy on the nucleation rate, thereby enhancing the formation of a metastable polymorph. This builds upon the understanding that low surface energies can reduce the barriers to nucleate metastable phases[3], which has been leveraged to synthesize binary oxides[6] and metal chalcogenides[7] from solutions. For example, nanosized $TiO_2$ is a well-studied system where the metastable anatase polymorph forms prior to the rutile one, despite the latter phase having a lower bulk free energy[8]. Similar observations have also been made regarding the crystallization of metastable polymorphs from melts[9,10], vapors[11], and amorphous media[12,13]. The current approach not only extends these ideas to solid-state synthesis, but also clarifies the quantitative role of reaction energy in dictating which polymorphs initially form. Because the reaction energy is set by the choice of precursors, it enables one to deliberately plan and control which polymorph is obtained, creating novel opportunities to synthesize metastable materials using a scalable, solid-state route.

$LiTiOPO_4$ (LTOPO) is used as a prototype system to demonstrate the validity of our framework. Two polymorphs have been reported for this compound, but the factors that govern the formation of each are yet undetermined. Density functional theory (DFT) calculations performed in this work reveal that the metastable polymorph of LTOPO has a lower surface energy than the ground state, and as such, may nucleate first at small particle size. We show that by using precursors with a large reaction energy to form LTOPO, the critical radius for nucleation is kept small enough to favor the metastable polymorph, whose formation is confirmed with *in situ* X-ray diffraction (XRD). In contrast, the use of precursors that form low-energy reaction intermediates require larger critical nuclei to form LTOPO, which we find leads to the formation of its stable polymorph. These findings support the hypothesis that reaction energy and its change along the synthesis path dictates the influence of surface energy on solid-state reaction outcomes, for which rational precursor selection can enable the targeted synthesis of metastable materials.

**Results**

**Thermodynamics of nucleation-controlled polymorph selection**

In a process controlled by nucleation, the selectivity of competing polymorphs can be assessed by comparing their nucleation rates. According to classical nucleation theory[14], the rate of nucleation ($Q$) for a given phase is related to its surface energy ($\gamma$) and the bulk free energy change ($\Delta G$) associated with its formation by the equation:

$$Q = A\exp\left(-\frac{16\pi\gamma^3}{3n^2 k_B T (\Delta G)^2}\right) \quad (1)$$

where $n$ is the number of atoms per unit volume, $T$ is temperature, $k_B$ is Boltzmann's constant, and $A$ is a pre-factor. According to Langer theory, $A$ is a product of the dynamic pre-factor $\kappa$, which is related to the growth rate of critical clusters, and the statistical pre-factor $\Omega_0$, which provides a measure of the phase space volume available for the nucleation[15].

In the context of solid-state synthesis, we refer to the bulk free energy change as the *reaction energy* ($\Delta G_{\text{rxn}}$). Whereas the surface energy of a phase is relatively constant in a given medium, assuming its nucleation is homogeneous, the reaction energy can be varied by modifying the reagents from which the phase forms. To illustrate the effects of surface and reaction energies, **Fig. 1a** plots several boundaries where the nucleation rate of a stable polymorph (*i*) is equal to that of a metastable polymorph (*j*) at the same composition. Each boundary represents a specific value of the bulk energy difference between the two polymorphs ($\Delta G_{i \to j}$) and is plotted as a function of two metrics: 1) the difference between the surface energies of the two polymorphs ($\gamma_i - \gamma_j$), and 2) the reaction energy to form the stable polymorph ($\Delta G_{\text{rxn}}$). Note that we assume $\gamma_j < \gamma_i$ in our analysis, otherwise polymorph *j* can never achieve an energetic advantage during nucleation.

In **Fig. 1a**, the region to the upper left of each boundary represents the conditions where the metastable polymorph (*j*) nucleates faster than the stable one (*i*), signifying an opportunity window to form the metastable phase. The range of this window depends on the bulk energy difference between the polymorphs. For a small energy difference, such as 10 meV/atom, preferential nucleation of the metastable polymorph occurs for a wide range of reaction energies (typically $\Delta G_{rxn} < -20$ meV/atom) even if its surface energy is only slightly lower ($> 5$ meV/Å$^2$) than that of the stable polymorph. A larger polymorph energy difference requires more extreme conditions to favor the metastable phase. For example, to access a polymorph with energy ≈ 100 meV/atom above the stable phase, the surface energy of the metastable phase must be ~20 meV/Å$^2$ lower than that of the stable phase to make its nucleation plausible. At the same time, the reaction energy must be more negative than $-80$ meV/atom to ensure that the metastable polymorph nucleates prior to the stable one.

The trends shown in **Fig. 1a** suggest that reaction energy is an effective handle to control the selectivity between two competing polymorphs. **Fig. 1b** further illustrates this by plotting the *critical reaction energy*, $\Delta G_{rxn}^*$ (Supplementary Note 1), below which a metastable polymorph ($j$) nucleates faster than its stable counterpart ($i$), against the ratio of the polymorph surface energies ($\gamma_j/\gamma_i$) and their bulk energy difference ($\Delta G_{i \to j}$). The plot shows that when two competing polymorphs have similar bulk formation energies, *i.e.*, small $\Delta G_{i \to j}$, only a small reaction energy is required to preferentially nucleate a metastable phase with lower surface energy ($\gamma_j/\gamma_i < 1$). In contrast, when polymorphs have a large bulk energy difference but similar surface energies, larger reaction driving force are required to access the metastable phase (*e.g.*, $\Delta G_{rxn} < -200$ meV/atom). Such large reaction driving force are less common, as they require highly reactive precursors that directly contribute to the product's formation without creating stable intermediates that consume the thermodynamic driving force.

The scales of the variables shown in **Figs. 1a and 1b** were chosen based on historical data from the literature, such that these diagrams cover the range of conditions where metastable polymorph nucleation is reasonably accessible. In the well-studied cases of metastable binary metal oxides synthesized from solution, polymorph surface energy differences have been reported to reach ~ 150 meV/Å² (**Fig. 1c**)[16,17]. Larger differences may also be achieved through heterogenous nucleation on a surface that favors the metastable polymorph, *e.g.*, when the product shares structural similarities with the precursors[18,19] or the substrate on which it is grown[20].

Bulk reaction energies span a similarly wide range of values. **Fig. 1d** shows the distribution of reaction energies from 7,562 prior solid-state synthesis experiments[21,22], calculated at a common synthesis temperature of 500 °C (Methods). Approximately 65.8% of reactions have $\Delta G_{\text{rxn}} < -50$ meV/atom, whereas 37.5% have $\Delta G_{\text{rxn}} < -200$ meV/atom. These results highlight the availability of precursors with large reaction energies, which may provide access to synthesize metastable polymorphs with low surface energies. These reported reaction energies will be reduced in the event of intermediate phase formation, emphasizing the need for *in situ* characterization.

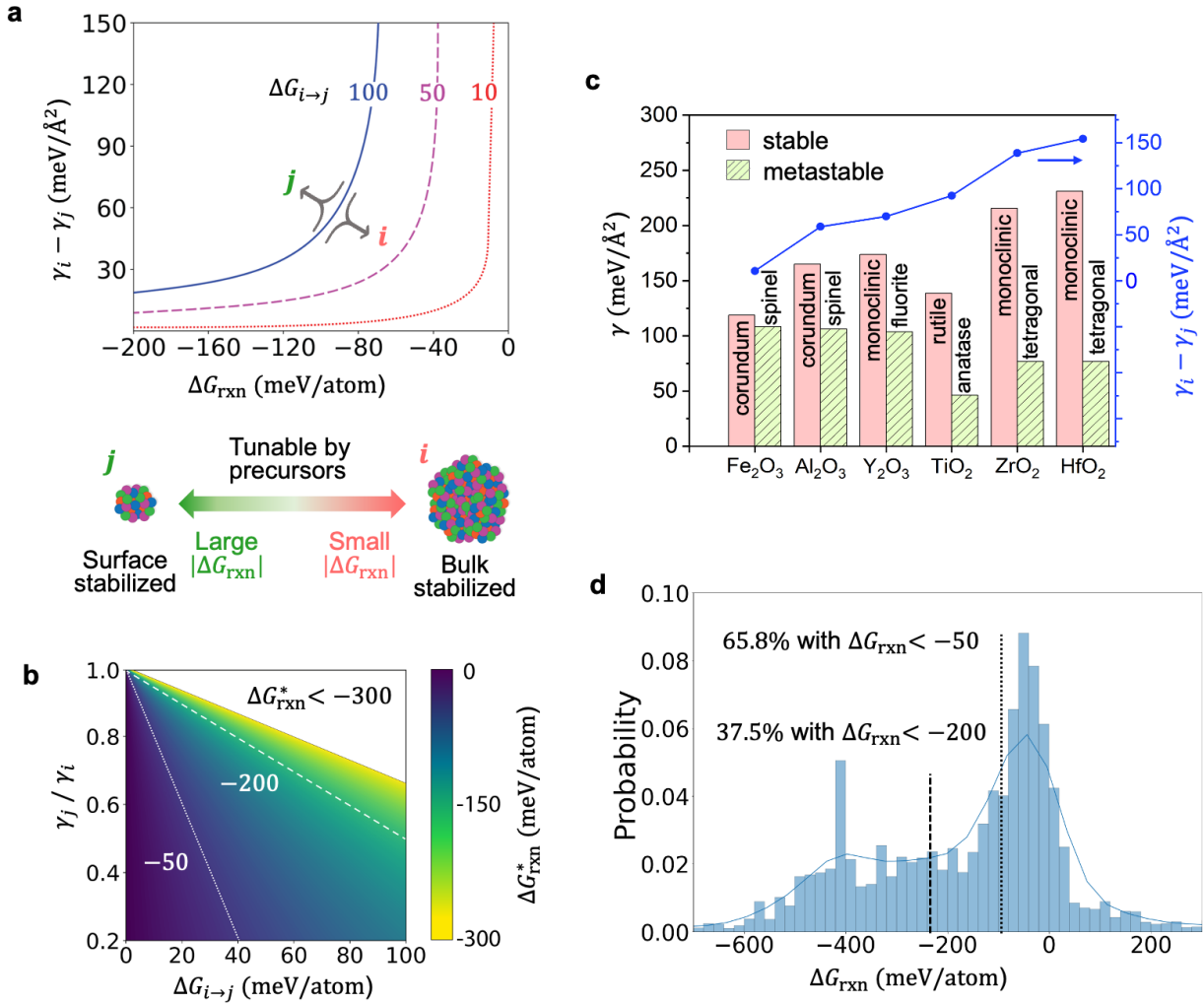

**Fig. 1: Opportunity windows for polymorph selection in solid-state synthesis. a**, Boundaries at which the nucleation rates of a stable polymorph (*i*) and metastable polymorph (*j*) with the same composition are equal. Axes denote the bulk reaction energy ($\Delta G_{rxn}$) and surface energy difference ($\gamma_i - \gamma_j$). Each curve is calculated for a distinct polymorph energy difference ($\Delta G_{i \to j}$) given in meV/atom on the graph. Schematic at the bottom shows the polymorph selectivity influenced by the bulk reaction energies that is tunable by precursor selection. **b**, A contour plot showing the critical reaction energy ($\Delta G_{rxn}^*$) required for preferential nucleation of a metastable polymorph at various $\Delta G_{i \to j}$ and $\gamma_j/\gamma_i$. **c**, Reported calorimetry-measured surface energies of anhydrous binary metal oxides $Fe_2O_3$[23], $Al_2O_3$[24], $Y_2O_3$[25], $TiO_2$[26], $ZrO_2$[27], and $HfO_2$[16,17]. **d**, Reaction energy distribution of solid-state reactions reported in the literature using a text-mined dataset[21,22] and energies calculated at 500 °C.

**LiTiOPO₄ polymorphs**

We test the validity of our framework describing polymorph selectivity in the synthesis of LiTiOPO$_4$ (LTOPO) which is known to form in either an orthorhombic (*o*-LTOPO) or a triclinic (*t*-LTOPO) polymorph (**Fig. 2a**). DFT calculations reveal that *o*-LTOPO is the ground state, whereas *t*-LTOPO is metastable with 12 meV/atom higher energy at 0 K. Nevertheless, both polymorphs have been observed experimentally, although the factors that dictate the selectivity of each polymorph were yet unclear. Previous studies have shown that *t*-LTOPO forms prior to *o*-LTOPO at low temperatures when using a solid-state route[28], whereas an opposite relation to temperature was observed during cooling-crystallization experiments[29], suggesting that temperature by itself does not dictate the polymorph selectivity. Our investigation of temperature effects based on vibrational entropy (**Fig. 2b**) confirms that the metastability of *t*-LTOPO remains unchanged throughout 0-1490 K, with the energy difference $\Delta G_{o \to t}$ increasing from 12 meV/atom to 21 meV/atom as the temperature rises.

The persistent metastability of *t*-LTOPO in its bulk form suggests that its experimental formation may be related to its nucleation kinetics at small particle size, where surface energies become important. The more favorable surface energy of *t*-LTOPO was verified using DFT calculations on slabs representing the low-index Miller indices for each polymorph. In **Fig. 2c**, we display the equilibrium particle shapes of *o*- and *t*-LTOPO determined using the Wulff construction. Consistent with the principle outlined by Navrotsky[3,30], which states that metastable polymorphs often have lower surface energies than their stable counterparts, the net surface energy of *t*-LTOPO (44.85 meV/Å$^2$, or 0.717 J/m$^2$) is lower than that of *o*-LTOPO (58.65 meV/Å$^2$, or 0.938 J/m$^2$). The enhanced stability of the metastable polymorph's surface can in largely be attributed to the low energy of its (100) facet, which constitutes 49% of the total surface area in the Wulff construction. As shown by the surface energy for each set of Miller indices in **Fig. 2d**, the (100) surface of *t*-LTOPO is more stable than any others of each polymorph, which we attribute to a reduced number of broken bonds along this termination (Supplementary Note 2).

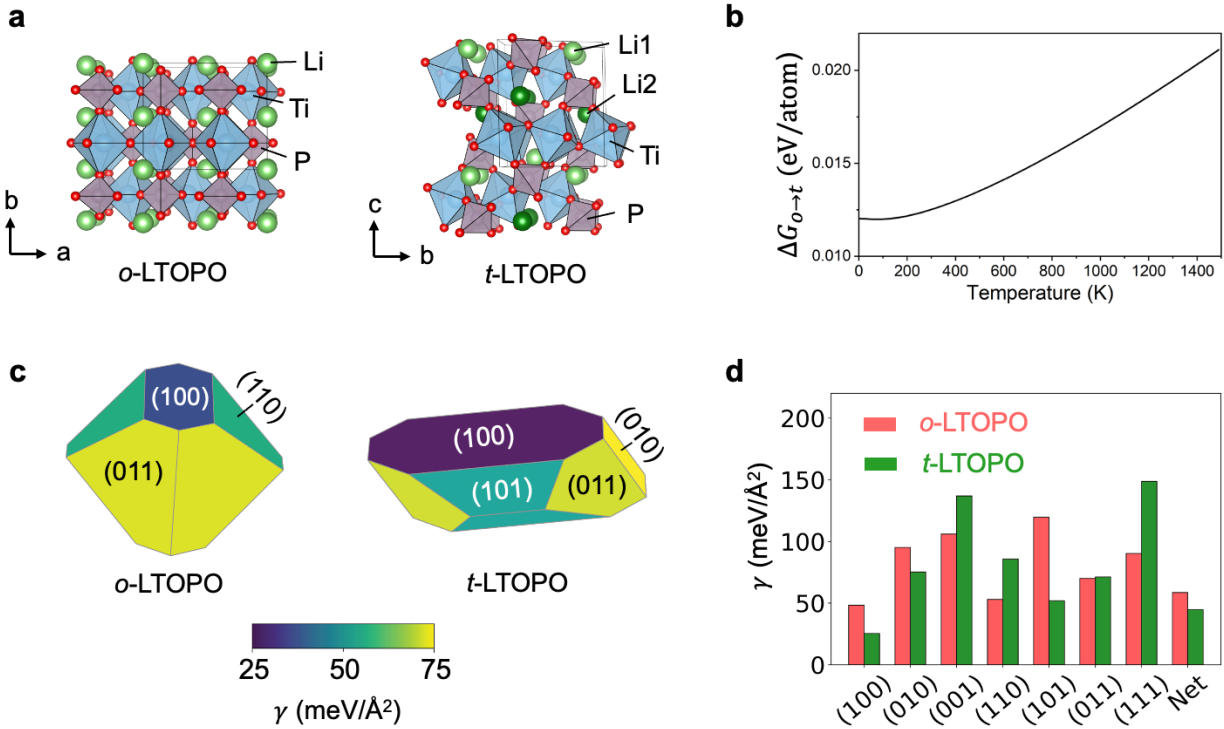

**Fig. 2: LiTiOPO₄ polymorph structures and energetics. a**, Crystal structures of the orthorhombic (*o*-LTOPO, $Pnma$) and triclinic (*t*-LTOPO, $P\bar{1}$) polymorphs for LiTiOPO$_4$. **b**, DFT-calculated free energy difference between *o*- and *t*-LTOPO from 0 K to 1490 K. **c**, Wulff constructions showing the equilibrium particle shape of each polymorph. **d**, DFT-calculated surface energies of all low-index surfaces comprising the Wulff construction of each polymorph. The net surface energies are also shown as the rightmost bars on the graph, revealing that *t*-LTOPO has a more stable surface than *o*-LTOPO.

***In situ* characterization of LiTiOPO₄ synthesis**

The stabilization of *t*-LTOPO by its low surface energy makes it an excellent candidate to probe the factors that influence polymorph selectivity. To this end, we performed solid-state synthesis experiments targeting LTOPO and monitored their phase evolution with *in situ* synchrotron XRD. Two precursor sets were investigated, differing only by their phosphate source – $P_2O_5$ versus $NH_4H_2PO_4$ – while using $Li_2CO_3$ and $TiO_2$ as the Li and Ti sources. Both precursor mixtures were ball milled at 450 rpm for 20 h to ensure intimate mixing. *In situ* synchrotron XRD measurements were then carried out on each mixture (in air) while heating at a rate of 25 °C/min to 700 °C, followed by a 3 h hold at this temperature. The selected precursors and conditions were chosen based on previous work where LTOPO was synthesized[28,29,31-33]. The phosphate source was changed to vary the reactivity of the starting precursor mixture, which we will show has a significant effect on the resulting synthesis pathway and product selectivity.

**Precursor Set 1: $Li_2CO_3$ + $TiO_2$ + $P_2O_5$**

**Figure 3a** shows a heatmap of the XRD intensities measured from precursor *Set 1* as it was heated to 700 °C. At low temperature, the patterns show only a few well-defined peaks that can be attributed to $TiO_2$. The other precursors become amorphous after ball milling, as evidenced by diffuse scattering in the XRD patterns (Supplementary Fig. 1). Near 500 °C, several peaks associated with *t*-LTOPO appear and continue to grow upon further heating, at the expense of the $TiO_2$ precursors whose signal decays between 500-700 °C. The weight fraction of each phase is plotted as a function of temperature in **Fig. 3b**. Since the sample was mostly amorphous at low temperature, we set the initial weight fraction of each phase ($Li_2O$, $TiO_2$, and $P_2O_5$) to its expected value based on the starting precursor stoichiometry. Rapid *t*-LTOPO growth from 500-550 °C is followed by slower growth upon further heating to 650 °C as the precursors are completely consumed and *t*-LTOPO becomes phase pure. This metastable polymorph remains present until 700 °C, at which point new peaks associated with *o*-LTOPO appear. The stable polymorph *o*-LTOPO continues to grow until it becomes the only remaining phase.

**Precursor Set 2: $Li_2CO_3$ + $TiO_2$ + $NH_4H_2PO_4$**

*In situ* synchrotron XRD measurements performed on precursor *Set 2* (**Fig. 3c**) reveal that $Li_2CO_3$ and $NH_4H_2PO_4$ reacted in the ball milling step, as evidenced by the appearance of $Li_3PO_4$ at low temperature. The weight fraction of each phase is plotted as a function of temperature in **Fig. 3d**. Upon heating, the partially reacted mixture of $TiO_2$ and $Li_3PO_4$ proceeds through a notably different reaction pathway than *Set 1*. Both *t*-LTOPO and $LiTi_2(PO_4)_3$ form as intermediates at 500 °C, consuming $TiO_2$ and $Li_3PO_4$ (**Fig. 3c**). $LiTi_2(PO_4)_3$ then contributes to the formation of *o*-LTOPO at 600 °C by reacting with leftover $Li_3PO_4$ and $TiO_2$. Further heating to 700 °C causes a phase transition from *t*-LTOPO to *o*-LTOPO, as was observed in the reaction pathway of *Set 1*.

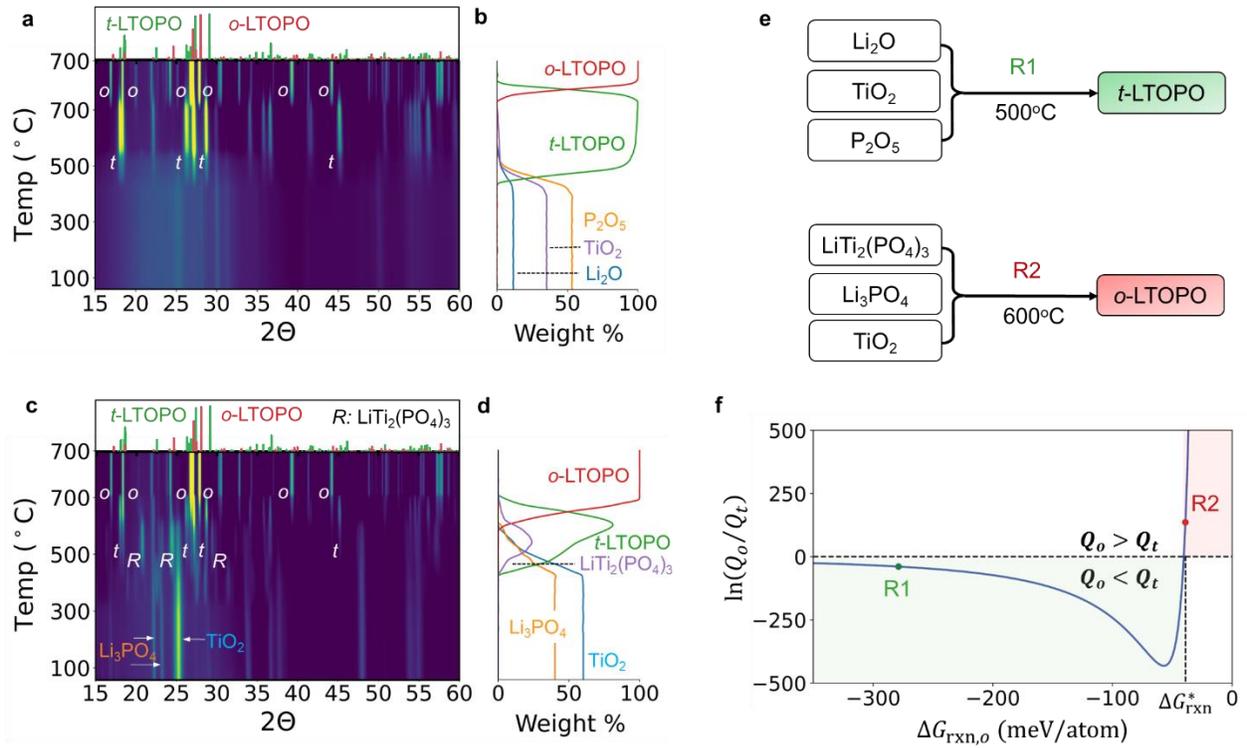

**Fig. 3: Phase evolution and polymorph selection during solid-state synthesis of LiTiOPO$_4$.** **a,c**, *In situ* synchrotron XRD patterns (2θ converted to Cu K$_α$) measured during heating to 700 °C, followed by a 3-hour hold, using starting precursors of **a**, (Set 1) Li$_2$CO$_3$, TiO$_2$, and P$_2$O$_5$ or **c**, (Set 2) Li$_2$CO$_3$, TiO$_2$, and NH$_4$H$_2$PO$_4$. **b,d**, Phase fraction evolution estimated from the peak intensity in **a** and **c**, respectively. Amounts of amorphous phases were calibrated based on the starting materials. **e**, Reaction pathways associated with the nucleation of *t*-LTOPO (R1) observed in **a** and nucleation of *o*-LTOPO (R2) observed in **c**. **f**, Relative polymorph nucleation rates between *o*-LTOPO and *t*-LTOPO as a function of reaction energy (Δ$G_{\text{rxn},o}$). The minimum thermodynamic driving force required to form the metastable *t*-LTOPO is denoted by Δ$G_{\text{rxn}}^*$. The two points (R1, R2) along the curve represent different reactant combinations that led to the initial formation of LTOPO.

***Ex situ*** **characterization of LiTiOPO$_4$ synthesis**

The reaction pathway followed by *Set 1* reveals a temperature window where the metastable *t*-LTOPO polymorph forms without impurities (500-700 °C), presenting a viable route to synthesize this phase. A separate synthesis procedure was performed by heating *Set 1* to 500 °C and holding for 1 h, after which *ex situ* XRD confirmed the presence of *t*-LTOPO without any detectable impurities (Supplementary Fig. 2). Furthermore, this phase remains unchanged even after holding the sample at 500 °C for 12 h, suggesting that the phase transformation of LTOPO polymorphs is more strongly dependent on temperature rather than time.

In contrast to *Set 1*, the combination of precursors in *Set 2* does not provide an effective route to synthesize the metastable *t*-LTOPO, as it never appears without any impurities in the reaction pathway. *Ex situ* XRD on a sample that was made by heating *Set 2* to 500 ºC (Supplementary Fig. 2) reveals a mixture of *t*-LTOPO and LiTi$_2$(PO$_4$)$_3$. This distinct path forms the stable *o*-LTOPO directly from intermediates, rather than from *t*-LTOPO. Holding the sample at 500 ºC for 12 h did not lead to any noticeable changes to the XRD pattern, which suggests that the transformation from LiTi$_2$(PO$_4$)$_3$ to *o*-LTOPO only occurs at higher temperatures.

**Effect of reaction energy on LiTiOPO$_4$ selectivity**

Based on the phase evolution observed through *in situ* XRD measurements, we identify two reactions that resulted in the formation of a LTOPO polymorph (**Fig. 3e**):

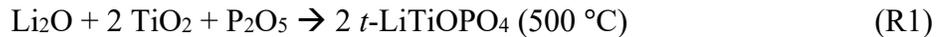

$$\text{Li}_2\text{O} + 2\ \text{TiO}_2 + \text{P}_2\text{O}_5 \rightarrow 2\ t\text{-LiTiOPO}_4\ (500\ °\text{C}) \qquad (\text{R1})$$

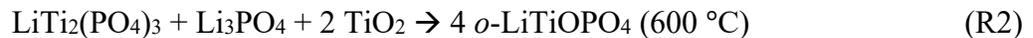

$$\text{LiTi}_2(\text{PO}_4)_3 + \text{Li}_3\text{PO}_4 + 2\ \text{TiO}_2 \rightarrow 4\ o\text{-LiTiOPO}_4\ (600\ °\text{C}) \qquad (\text{R2})$$

Despite occurring at similar temperatures, R1 forms the metastable polymorph (*t*-LTOPO) while R2 forms the ground state (*o*-LTOPO). The key difference between R1 and R2 lies in their reaction energies. DFT calculations indicate that R1 has a much larger reaction energy (-279 meV/atom) than R2 (-40 meV/atom). Our framework suggests that R1 should therefore favor the nucleation of whichever polymorph has the lower surface energy. Indeed, calculation of the relative nucleation rates (Supplementary Note 3) suggests that *t*-LTOPO nucleates ~$10^{17}$ times faster than *o*-LTOPO when preceded by the reagents in R1. In contrast, the smaller reaction energy associated with R2 favors larger particles with more stable bulk energies, and our calculation of the relative nucleation rates indicate that *o*-LTOPO nucleates ~$10^{59}$ times faster than *t*-LTOPO in this case.

**Figure 3f** gives the logarithm of the relative nucleation rates of *o*-LTOPO ($Q_o$) and *t*-LTOPO ($Q_t$) as a function of a wide range of reaction energies. The resulting curve highlights wo distinct regimes for each polymorph to preferentially nucleate, separated by the *critical reaction energy* ($\Delta G_{rxn}^*$), which we define as the minimum thermodynamic driving force required to nucleate the metastable polymorph faster than the stable one (*i.e.*, $Q_t > Q_o$). Because the calculated reaction energy for R1 is larger than the critical reaction energy, it falls within the regime where *t*-LTOPO is expected to nucleate first (green). In contrast, the smaller reaction energy of R2 does not satisfy the critical reaction energy requirement, and therefore it favors nucleation of the stable polymorph, *o*-LTOPO (red).

**Discussion**

The targeted synthesis of metastable polymorphs bolsters current approaches to materials design by enabling access to a vastly enlarged space beyond thermodynamically stable materials. The formation of metastable phases with lower surface energy has long been studied in solution-based methods, where the small particle size enables surface energy to dictate reaction outcomes. Related work in thin films and amorphous media has demonstrated that metastable polymorphs can also be accessed by modifying the surface energy through structural templating or epitaxial growth[18-20], *i.e.*, by engineering the rate of heterogenous nucleation. We have shown in this work that metastable polymorphs are also accessible in traditional solid-state synthesis where controlled modification of the surface energy is typically considered to be more challenging. Our work demonstrates that the reaction energy, easily modifiable by changing precursors, can be used as an additional handle to control the relative nucleation rates of competing polymorphs. Larger reaction energies effectively reduce the critical radius required for nucleation, thereby increasing the ratio of surface area to bulk volume in the corresponding nuclei (**Fig. 1a**). As such, nucleation events driven by large reaction energies tend to favor the formation of products with low surface energies. This principle was successfully applied to synthesize a metastable polymorph of $LiTiOPO_4$, whose formation is made possible by 1) its surface energy being lower (more stable) than that of the ground state, and 2) the use of precursors that maintain a large reaction energy to form it.

The framework for polymorph selectivity presented here operates under the assumption of homogeneous nucleation, rather than heterogenous, and the calculation of nucleation rates is based on classical nucleation theory. In reality, the nucleation of phases formed *via* solid-state reactions generally follow more complex mechanisms. For example, nuclei tend to form at the interfaces between different particles as well as at the surface of the sample container, often leading to reduction in the nucleation barrier. Furthermore, previous work has shown that nucleation may proceed through an amorphous intermediate, which also lowers the barrier for nucleation[34]. However, accounting for such effects in a quantitative manner is difficult using first-principles calculations, and we believe that our methodology can still be used to provide approximate guidance for polymorph selectivity in solid-state synthesis as heterogeneous nucleation only influences a fraction of the nucleating solid's surface. In general, the guidance provided by our framework is mostly likely to hold true in extreme cases (*e.g.*, large surface energy differences), as demonstrated for LiTiOPO$_4$.

The data presented in **Fig. 1** suggests that there exists a wide range of conditions under which surface-stabilized metastable polymorphs may be obtained through solid-state synthesis. A key requirement for this task is the selection of optimal precursors that not only start with a large reaction energy to form the desired target, but also maintain it should any intermediates form. To this end, our framework for polymorph selectivity may benefit from integration with existing techniques for the design[35-37] and optimization[38] of reaction pathways. Furthermore, coupling these techniques with *in situ* characterization[19,39] would provide experimental validation of the proposed reaction pathways, enabling a complete understanding of the factors that dictate the synthesis of metastable polymorphs in the solid state.

**Conclusion**

We have introduced a general scheme for polymorph selectivity, where the preferential nucleation of a certain phase is dictated by three variables: bulk energy difference, surface energy difference, and reaction energy. Using LiTiOPO$_4$ (LTOPO) as an example, we demonstrate that reaction energy provides an effective handle to control polymorph selectivity in solid-state synthesis. *In situ* XRD measurements revealed that the metastable polymorph of LTOPO forms prior to the ground state during solid-state synthesis, but its yield is strongly dependent on the choice of precursors. DFT calculations showed that the metastable phase is stabilized by a low surface energy, which becomes a dominant factor in its nucleation rate when the reaction energy to form it is sufficiently large. These findings suggest new opportunities to synthesize metastable materials using a solid-state route, where highly reactive precursors can be selected to maximize the reaction driving force and therefore favor the nucleation of metastable phases with low surface energies.

**Methods**

**Synthesis and Characterizations**

Li$_2$CO$_3$, TiO$_2$, P$_2$O$_5$, and NH$_4$H$_2$PO$_4$ were purchased from Sigma Aldrich and used directly for the synthesis of LTOPO. To prepare precursor mixtures, chemicals were weighed and loaded into a ZrO$_2$-lined jar in an Ar-filled glovebox. 10 wt% excess Li$_2$CO$_3$ was added to compensate for any Li loss during high-energy ball milling or high-temperature treatment. The powder was mixed using ten ZrO$_2$ grinding balls with 10 mm diameter and milled at 450 rpm for 20 h in a Retsch PM 400 planetary ball mill. After ball-milling, the powder was scraped from the jar and then pelletized. For *in situ* studies, the pellets were broken to small pieces to fit into a quartz capillary. For *ex situ* studies, heat treatment on the pellets was conducted in a box furnace in the air. After heating, the pellets were removed from the furnace and fast cooled in the air. Pellets were pulverized to fine powder using a mortar and pestle. *Ex situ* XRD was performed using a Rigaku Miniflex 600 diffractometer with Cu K$_\alpha$ radiation.

***In Situ* Synchrotron XRD**

*In situ* synchrotron XRD was performed at beamline 11-ID-B at the Advanced Photon Source (APS) of Argonne National Laboratory with a constant wavelength of 0.2115 Å. Samples were loaded into 1.1 mm quartz capillaries and mounted in a flow cell optimized for the collection of diffraction data in transmission geometry[40]. The flow cell was mounted at the beamline on an x-y stage for ease of alignment. Samples were heated with compact resistive heating elements to temperatures up to 700 °C at a ramp rate of 25 °C/min under air without any gas flow. Diffraction data were acquired every 15 s on an amorphous silicon-based area detector (PerkinElmer XRD1621) positioned at a nominal distance of 1000 mm from the sample. Calibration of the beam center, sample-detector distance, rotation, and tilt angle were performed in GSAS-II using a CeO$_2$ standard[41]. Reduction of the 2-dimensional images to 1-dimensional patterns was performed in GSAS-II.

**Bulk Free Energies**

For all solid phases studied in this work, bulk free energies were calculated using DFT calculations performed with the Vienna *ab initio* simulation package (VASP)[42-45]. Starting structures were taken from Materials Project[46] and relaxed using the projector augmented wave (PAW) method with the strongly constrained and appropriately normed (SCAN) functional[47]. A cutoff energy of 600 eV was imposed on the plane wave basis sets. For each structure, the Brillouin zone was sampled with Gaussian smearing (0.05 eV width) on a Γ-centered mesh containing at least 25 k-points per Å$^{-1}$. Unit cells and atomic positions were relaxed until all forces were less than 10$^{-2}$ eV/Å. On the final structures, electronic optimization was performed using the tetrahedron method with Blöchl corrections[48] and a convergence criterion of 10$^{-6}$ eV.

To account for finite temperature effects, vibrational entropies were computed for both LTOPO polymorphs through application of the quasi-harmonic approximation (QHA) based on density functional perturbation theory (DFPT), as implemented in the Phonopy package[49]. Supercells of size $2 \times 2 \times 1$ were prepared based on the DFT-relaxed structure of each polymorph. To apply QHA, the volumes ($V$) of these supercells were expanded and compressed to form nine distinct structures with linear strains $\Delta\varepsilon \in \{-3\%, -2\%, -1\%, -0.5\%, 0\%, 0.5\%, 1\%, 2\%,$ 3%}. DFPT calculations were performed using the Perdew–Burke–Ernzerhof generalized gradient approximation (GGA) [ref: Generalized gradient[50] using the projector augmented wave method [51]. An energy cutoff of 520 eV was and a stricter energy convergence criterion of $10^{-8}$ eV was used for the DFPT calculations. After obtaining the vibrational entropies ($S$), the Gibbs free energy ($G$) was calculated as a function of temperature ($T$) for each LTOPO polymorph:

$$G = H - TS$$

where $H$ is the enthalpy of each phase, approximated by the DFT-calculated energy. For all other phases considered in this work, temperature-dependent Gibbs free energies were estimated using the machine-learned descriptor developed by Bartel *et al.*[52], which can closely approximate the vibrational entropies in solid phases.

**Surface Energies**

For each polymorph of LTOPO, the surface energy was determined by performing DFT calculations on surface slabs generated using the efficient creation and convergence scheme[53], as implemented in the Python Materials Genomic (Pymatgen) package[54]. Only low-index surfaces were considered, including Miller indices ($hkl$) with $h, k, l \in \{\bar{1}, 0, 1\}$. Slabs were generated with a thickness of at least 10 Å and a 15 Å vacuum. Atomic positions within each slab were relaxed while keeping the unit cell fixed to maintain the interlayer vacuum. The parameters used for the DFT calculations performed here were the same as those used for the bulk free energy calculations, except for the k-point mesh, where only the Γ point was sampled along the direction normal to the surface. From the final energies of the relaxed slabs ($E_{slab}$), surface energies ($\gamma$) were calculated as follows:

$$\gamma = \frac{1}{2A}(E_{slab} - NE_{bulk})$$

where $A$ is the surface area of the slab, $N$ is the number of atoms it contains, and $E_{bulk}$ is the normalized (per atom) energy of the bulk phase. The Wulff construction was used to determine the equilibrium particle shape for each polymorph of LTOPO, from which total surface energies were calculated.

**Text Mining Dataset**

To investigate distribution of reaction energies in solid-state synthesis experiments, we extracted the information associated with 7,562 solid-state reactions from a previously reported dataset that was formed by text-mining the scientific literature[21,22]. For each reaction, the difference in the Gibbs free energies of the product(s) and the precursor(s) were calculated using thermochemical data from the Materials Project (MP)[46]. Each chemical formula was mapped to the lowest-energy structure or a linear combination of them available in the MP for that composition. Since all MP energies are calculated at 0 K, we approximate the finite-temperature Gibbs free energy of each phase at 500 °C using the machine-learned descriptor developed by Bartel *et al.*[52]. For gaseous species such as $O_2$ and $CO_2$, the temperature dependent enthalpy and entropy were taken from the FREED[55] and NIST[56] experimental databases. For materials containing $CO_3^{2-}$ anions, an empirical correction of -1.2485 eV/$CO_3$ was applied to compensate for systematic errors in density functional theory. This value was calibrated based on experimental enthalpies of common carbonates[21].

# References

1. Aykol, M., Dwaraknath, S. S., Sun, W. & Persson, K. A. Thermodynamic limit for synthesis of metastable inorganic materials. *Science Advances* **4**, eaaq0148 (2018). https://doi.org:doi:10.1126/sciadv.aaq0148
2. Sun, W. *et al.* The thermodynamic scale of inorganic crystalline metastability. *Science Advances* **2**, e1600225 (2016). https://doi.org:doi:10.1126/sciadv.1600225
3. Navrotsky, A. Energetic clues to pathways to biomineralization: Precursors, clusters, and nanoparticles. *Proceedings of the National Academy of Sciences* **101**, 12096-12101 (2004). https://doi.org:doi:10.1073/pnas.0404778101
4. Cordova, D. L. M. & Johnson, D. C. Synthesis of Metastable Inorganic Solids with Extended Structures. *ChemPhysChem* **21**, 1345-1368 (2020). https://doi.org:https://doi.org/10.1002/cphc.202000199
5. Heinselman, K. N., Lany, S., Perkins, J. D., Talley, K. R. & Zakutayev, A. Thin Film Synthesis of Semiconductors in the Mg–Sb–N Materials System. *Chemistry of Materials* **31**, 8717-8724 (2019). https://doi.org:10.1021/acs.chemmater.9b02380
6. Navrotsky, A. Energetics of nanoparticle oxides: interplay between surface energy and polymorphism†. *Geochemical Transactions* **4**, 34 (2003). https://doi.org:10.1186/1467-4866-4-34
7. Washington, A. L. *et al.* Ostwald's Rule of Stages and Its Role in CdSe Quantum Dot Crystallization. *Journal of the American Chemical Society* **134**, 17046-17052 (2012). https://doi.org:10.1021/ja302964e
8. Hanaor, D. A. H. & Sorrell, C. C. Review of the anatase to rutile phase transformation. *Journal of Materials Science* **46**, 855-874 (2011). https://doi.org:10.1007/s10853-010-5113-0
9. Volkmann, T., Herlach, D. M. & Löser, W. Nucleation and phase selection in undercooled Fe-Cr-Ni melts: Part I. Theoretical analysis of nucleation behavior. *Metallurgical and Materials Transactions A* **28**, 453-460 (1997). https://doi.org:10.1007/s11661-997-0146-y
10. Oyelaran, O., Novet, T., Johnson, C. D. & Johnson, D. C. Controlling Solid-State Reaction Pathways: Composition Dependence in the Nucleation Energy of InSe. *Journal of the American Chemical Society* **118**, 2422-2426 (1996). https://doi.org:10.1021/ja953560k
11. Cao, K. *et al.* Atomic mechanism of metal crystal nucleus formation in a single-walled carbon nanotube. *Nature Chemistry* **12**, 921-928 (2020). https://doi.org:10.1038/s41557-020-0538-9
12. Chung, S.-Y., Kim, Y.-M., Kim, J.-G. & Kim, Y.-J. Multiphase transformation and Ostwald's rule of stages during crystallization of a metal phosphate. *Nature Physics* **5**, 68-73 (2009). https://doi.org:10.1038/nphys1148
13. Stone, K. H. *et al.* Influence of amorphous structure on polymorphism in vanadia. *APL Materials* **4**, 076103 (2016). https://doi.org:10.1063/1.4958674
14. Mullin, J. W. in *Crystallization (Fourth Edition)* (ed J. W. Mullin) 181-215 (Butterworth-Heinemann, 2001).
15. Langer, J. S. Metastable states. *Physica* **73**, 61-72 (1974). https://doi.org:https://doi.org/10.1016/0031-8914(74)90226-2
16. Zhou, W. *et al.* Hafnia: Energetics of thin films and nanoparticles. *Journal of Applied Physics* **107**, 123514 (2010). https://doi.org:10.1063/1.3435317

17  Ushakov, S. V. & Navrotsky, A. Direct measurements of water adsorption enthalpy on hafnia and zirconia. *Applied Physics Letters* **87**, 164103 (2005). https://doi.org:10.1063/1.2108113

18  Shannon, R. D. & Rossi, R. C. Definition of Topotaxy. *Nature* **202**, 1000-1001 (1964). https://doi.org:10.1038/2021000a0

19  Bai, J. *et al.* Kinetic Pathways Templated by Low-temperature Intermediates during Solid-state Synthesis of Layered Oxides. *Chemistry of Materials* (2020). https://doi.org:10.1021/acs.chemmater.0c02568

20  Nagurney, A. B., Caddick, M. J., Pattison, D. R. M. & Michel, F. M. Preferred orientations of garnet porphyroblasts reveal previously cryptic templating during nucleation. *Scientific Reports* **11**, 6869 (2021). https://doi.org:10.1038/s41598-021-85525-7

21  Huo, H. *et al.* Machine-Learning Rationalization and Prediction of Solid-State Synthesis Conditions. *Chemistry of Materials* (2022). https://doi.org:10.1021/acs.chemmater.2c01293

22  Kononova, O. *et al.* Text-mined dataset of inorganic materials synthesis recipes. *Scientific Data* **6**, 203 (2019). https://doi.org:10.1038/s41597-019-0224-1

23  Navrotsky, A., Mazeina, L. & Majzlan, J. Size-Driven Structural and Thermodynamic Complexity in Iron Oxides. *Science* **319**, 1635-1638 (2008). https://doi.org:doi:10.1126/science.1148614

24  McHale, J. M., Auroux, A., Perrotta, A. J. & Navrotsky, A. Surface Energies and Thermodynamic Phase Stability in Nanocrystalline Aluminas. *Science* **277**, 788-791 (1997). https://doi.org:doi:10.1126/science.277.5327.788

25  Zhang, P. *et al.* Energetics of Cubic and Monoclinic Yttrium Oxide Polymorphs: Phase Transitions, Surface Enthalpies, and Stability at the Nanoscale. *The Journal of Physical Chemistry C* **112**, 932-938 (2008). https://doi.org:10.1021/jp7102337

26  Levchenko, A. A., Li, G., Boerio-Goates, J., Woodfield, B. F. & Navrotsky, A. TiO2 Stability Landscape: Polymorphism, Surface Energy, and Bound Water Energetics. *Chemistry of Materials* **18**, 6324-6332 (2006). https://doi.org:10.1021/cm061183c

27  Radha, A. V., Bomati-Miguel, O., Ushakov, S. V., Navrotsky, A. & Tartaj, P. Surface Enthalpy, Enthalpy of Water Adsorption, and Phase Stability in Nanocrystalline Monoclinic Zirconia. *Journal of the American Ceramic Society* **92**, 133-140 (2009). https://doi.org:https://doi.org/10.1111/j.1551-2916.2008.02796.x

28  Morimoto, H. *et al.* Charge/discharge Behavior of Triclinic $LiTiOPO_4$ Anode Materials for Lithium Secondary Batteries. *Electrochemistry* **84**, 878-881 (2016). https://doi.org:10.5796/electrochemistry.84.878

29  I. N. Geifman, N. G. F., P. G. Nagornyi, L. D. Yun, M. V. Rotenfel'd. Crystal structure and V4+ EPR of the Li-Ti double oxyorthophosphate α-LiTiOPO4. *Kristallografia* **38**, 6 (1993).

30  Navrotsky, A. Nanoscale Effects on Thermodynamics and Phase Equilibria in Oxide Systems. *ChemPhysChem* **12**, 2207-2215 (2011). https://doi.org:https://doi.org/10.1002/cphc.201100129

31  Hofmann, P. *et al.* Structural analysis and electrical characterization of cation-substituted lithium ion conductors $Li_{1-x}Ti_{1-x}M_xOPO_4$ (M = Nb, Ta, Sb). *Solid State Ionics* **319**, 170-179 (2018). https://doi.org:https://doi.org/10.1016/j.ssi.2018.01.049

| | |
|---|---|
| 32 | Fu, Y. *et al.* A new insight into the LiTiOPO4 as an anode material for lithium ion batteries. *Electrochimica Acta* **185**, 211-217 (2015). https://doi.org:https://doi.org/10.1016/j.electacta.2015.10.124 |
| 33 | Chakir, M., El Jazouli, A., Chaminade, J. P., Bouree, F. & de Waal, D. New Process of Preparation, X-Ray Characterization, Structure and Vibrational Studies of a Solid Solution LiTiOAs1-xPxO4 (0 ≤ x ≤ 1). *ChemInform* **37** (2006). https://doi.org:https://doi.org/10.1002/chin.200616001 |
| 34 | Peng, Y. *et al.* Two-step nucleation mechanism in solid–solid phase transitions. *Nature Materials* **14**, 101-108 (2015). https://doi.org:10.1038/nmat4083 |
| 35 | McDermott, M. J., Dwaraknath, S. S. & Persson, K. A. A graph-based network for predicting chemical reaction pathways in solid-state materials synthesis. *Nat Commun* **12**, 3097 (2021). https://doi.org:10.1038/s41467-021-23339-x |
| 36 | Aykol, M., Montoya, J. H. & Hummelshøj, J. Rational Solid-State Synthesis Routes for Inorganic Materials. *Journal of the American Chemical Society* **143**, 9244-9259 (2021). https://doi.org:10.1021/jacs.1c04888 |
| 37 | Todd, P. K. *et al.* Selectivity in Yttrium Manganese Oxide Synthesis via Local Chemical Potentials in Hyperdimensional Phase Space. *Journal of the American Chemical Society* **143**, 15185-15194 (2021). https://doi.org:10.1021/jacs.1c06229 |
| 38 | Szymanski, N. J., Nevatia, P., Bartel, C. J., Zeng, Y. & Ceder, G. Autonomous decision making for solid-state synthesis of inorganic materials. arXiv:2304.09353 (2023). <https://ui.adsabs.harvard.edu/abs/2023arXiv230409353S>. |
| 39 | Bianchini, M. *et al.* The interplay between thermodynamics and kinetics in the solid-state synthesis of layered oxides. *Nature Materials* **19**, 1088-1095 (2020). https://doi.org:10.1038/s41563-020-0688-6 |
| 40 | Chupas, P. J. *et al.* A versatile sample-environment cell for non-ambient X-ray scattering experiments. *Journal of Applied Crystallography* **41**, 822-824 (2008). https://doi.org:doi:10.1107/S0021889808020165 |
| 41 | Toby, B. H. & Von Dreele, R. B. GSAS-II: the genesis of a modern open-source all purpose crystallography software package. *Journal of Applied Crystallography* **46**, 544-549 (2013). https://doi.org:doi:10.1107/S0021889813003531 |
| 42 | Kresse, G. & Furthmüller, J. Efficient iterative schemes for ab initio total-energy calculations using a plane-wave basis set. *Physical Review B* **54**, 11169-11186 (1996). https://doi.org:10.1103/PhysRevB.54.11169 |
| 43 | Kresse, G. & Furthmüller, J. Efficiency of ab-initio total energy calculations for metals and semiconductors using a plane-wave basis set. *Computational Materials Science* **6**, 15-50 (1996). https://doi.org:https://doi.org/10.1016/0927-0256(96)00008-0 |
| 44 | Kresse, G. & Hafner, J. Ab initio molecular-dynamics simulation of the liquid-metal--amorphous-semiconductor transition in germanium. *Physical Review B* **49**, 14251-14269 (1994). https://doi.org:10.1103/PhysRevB.49.14251 |
| 45 | Kresse, G. & Hafner, J. Ab initio molecular dynamics for liquid metals. *Phys Rev B Condens Matter* **47**, 558-561 (1993). https://doi.org:10.1103/physrevb.47.558 |
| 46 | Jain, A. *et al.* Commentary: The Materials Project: A materials genome approach to accelerating materials innovation. *APL Materials* **1**, 011002 (2013). https://doi.org:10.1063/1.4812323 |
| 47 | Sun, J., Ruzsinszky, A. & Perdew, J. P. Strongly Constrained and Appropriately Normed Semilocal Density Functional. *Physical Review Letters* **115**, 036402 (2015). |

48  Blöchl, P. E., Jepsen, O. & Andersen, O. K. Improved tetrahedron method for Brillouin-zone integrations. *Physical Review B* **49**, 16223-16233 (1994). https://doi.org:10.1103/PhysRevB.49.16223
49  Togo, A. & Tanaka, I. First principles phonon calculations in materials science. *Scripta Materiala* **108**, 1-5 (2015).
50  Perdew, J. P., Burke, K. & Ernzerhof, M. Generalized Gradient Approximation Made Simple. *Phys Rev Lett* **77**, 3865-3868 (1996). https://doi.org:10.1103/PhysRevLett.77.3865
51  Blöchl, P. E. Projector augmented-wave method. *Physical Review B* **50**, 17953-17979 (1994). https://doi.org:10.1103/PhysRevB.50.17953
52  Bartel, C. J. *et al.* Physical descriptor for the Gibbs energy of inorganic crystalline solids and temperature-dependent materials chemistry. *Nat Commun* **9**, 4168 (2018). https://doi.org:10.1038/s41467-018-06682-4
53  Sun, W. & Ceder, G. Efficient creation and convergence of surface slabs. *Surface Science* **617**, 53-59 (2013).
54  Ong, S. P. *et al.* Python Materials Genomics (pymatgen): A robust, open-source python library for materials analysis. *Computational Materials Science* **68**, 314-319 (2013).
55  *FREED-Thermodynamic Database*, <https://www.thermart.net/freed-thermodynamic-database/>
56  *NIST Chemistry WebBook*, <https://webbook.nist.gov/chemistry/>

**Acknowledgments**

This work was funded by the U.S. Department of Energy, Office of Science, Office of Basic Energy Sciences, Materials Sciences and Engineering Division under Contract No. DE-AC02-05-CH11231 (D2S2 program, KCD2S2). Computations were performed using the National Energy Research Scientific Computing Center (NERSC), a DOE Office of Science User Facility supported by the Office of Science and the U.S. Department of Energy under Contract No. DE-AC02-05CH11231. N.J.S. was supported in part by the National Science Foundation Graduate Research Fellowship under grant #1752814.

**Author contributions**

Y.Z., N.J.S., and G.C. planned this work. Y.Z. and N.J.S. formulated the theoretic framework. Y.Z. conducted synthesis and characterizations. N.J.S. calculated and analyzed the surface energy and reaction energy. T.H., and H.H., and C.J.B. extracted and analyzed the text-mining dataset. K.J. performed DFT calculations on the free energies of LiTiOPO$_4$. L.C.G. conducted in situ synchrotron XRD. C.J.B. contributed to the theoretical framework and assisted in surface energy calculations. B.O. contributed to the theorical framework and DFT calculations. Y.Z., N.J.S., and G.C. wrote the paper. All the authors contributed to the discussion and writing. G.C. supervised all aspects of the research.

**Competing interests**

The authors declare no competing interests.

# Supplementary Information

# Selective formation of metastable polymorphs in solid-state synthesis

Yan Zeng[+], Nathan J. Szymanski[+], Tanjin He, KyuJung Jun, Leighanne C. Gallington,

Haoyan Huo, Christopher J. Bartel, Bin Ouyang, Gerbrand Ceder[*]

Supplementary Notes 1-3

Supplementary Figs. 1-3

**Supplementary Note 1: Calculation of nucleation rates**

Based on classical nucleation theory[1], the rate of nucleation ($Q$) for a given phase is related to its surface energy ($\gamma$) and per-atom reaction energy ($\Delta G_{rxn}$) by the equation:

$$Q = A\exp\left(-\frac{16\pi\gamma^3}{3n^2 k_B T (\Delta G_{rxn})^2}\right) \tag{S1}$$

where $n$ is the number of atoms per unit volume, $T$ is temperature, $k_B$ is Boltzmann's constant, and $A$ is a pre-factor. According to Langer theory, $A$ is a product of the dynamic pre-factor $\kappa$, which is related to the growth rate of critical clusters, and the statistical pre-factor $\Omega_0$, which provides a measure of the phase space volume available for the nucleation[2]. The ratio between the nucleation rates of two species, $i$ and $j$, is given by:

$$\frac{Q_i}{Q_j} = \frac{A_i}{A_j}\exp\left[\frac{16\pi}{3k_B T}\left(\frac{(\gamma_j)^3}{n_j^2(\Delta G_{rxn,j})^2} - \frac{(\gamma_i)^3}{n_i^2(\Delta G_{rxn,i})^2}\right)\right] \tag{S2}$$

When $i$ and $j$ are two different polymorphs with equal compositions, their reaction energies are simply offset by their bulk free energy difference, $\Delta G_{i\rightarrow j}$. Assuming $j$ is the higher energy (metastable) phase, $\Delta G_{i\rightarrow j}$ will always be positive. Here, we also assume that $A_i \approx A_j$ as their difference will likely have a negligible influence on the relative nucleation rate of each polymorph, as compared with the stronger influence of the exponential term[3]. Under these assumptions, the ratio of the nucleation rates between two competing polymorphs can be estimated by:

$$\frac{Q_i}{Q_j} = \exp\left[\frac{16\pi}{3k_B T}\left(\frac{(\gamma_j)^3}{n_j^2(\Delta G_{rxn,i}+\Delta G_{i\rightarrow j})^2} - \frac{(\gamma_i)^3}{n_i^2(\Delta G_{rxn,i})^2}\right)\right] \tag{S3}$$

When the two polymorphs ($i$ and $j$) have equal nucleation rates and assuming their atomic densities are the same:

$$\Delta G_{rxn,i} = \frac{\Delta G_{i\rightarrow j}}{(\gamma_j/\gamma_i)^{3/2}-1} \tag{S4}$$

We denote the $\Delta G_{\text{rxn},i}$ under this circumstance as the *critical reaction energy* (minimum thermodynamic driving force) ($\Delta G_{\text{rxn}}^*$) required for preferential nucleation of a metastable polymorph.

**Supplementary Note 2: Analysis of surface energies of LiTiOPO$_4$**

Previous work has demonstrated that the surface energies of crystalline facets are often related to their associated densities of broken bonds[4,5]. We probe this relation for LTOPO by visualizing the (100) surface structure for each polymorph in Supplementary Fig. 3. In both structures, the PO$_4^{3-}$ polyhedra are completely preserved (*i.e.*, no broken P−O bonds). On the other hand, some Li−O and Ti−O bonds must be cleaved to form the (100) facet. As shown by the densities of broken bonds in Supplementary Fig. 3, both polymorphs of LTOPO share a similar number of broken Ti−O bonds along the (100) facet. These correspond to the loss of one coordinated O from each octahedral Ti complex within the surface structures. In contrast, the density of broken Li−O bonds differs greatly between polymorphs. *o*-LTOPO contains ~0.112 broken Li−O bonds per Å$^2$, whereas *t*-LTOPO contains only ~0.035 broken Li-O bonds per Å$^2$. This difference occurs in part because the Li ions within the bulk structure of *t*-LTOPO have a lower nearest-neighbor coordination (five O) than those in *t*-LTOPO (six O), and therefore less bonds must be cleaved to form a surface. Moreover, given the reduced connectivity of Li−O polyhedra in *t*-LTOPO as compared to those in *o*-LTOPO, the loss of O affects less coordinated Li in the former structure. These factors contribute to the low surface energy of the (100) facet in *t*-LTOPO and are therefore critical to its low nucleation barrier.

**Supplementary Note 3: Calculation of relative nucleation rate**

We use Eqn. S3 to assess the relative nucleation rates between *t*-LTOPO and *o*-LTOPO based on their calculated surface energies ($\gamma_t = 44.85$ meV/Å$^2$, $\gamma_o = 58.65$ meV/Å$^2$), bulk formation energy difference ($\Delta G_{o \to t} = 12$ meV/atom), atomic densities ($n_t = 0.089$ atom/Å$^3$, $n_o = 0.090$ atom/Å$^3$), and reaction energies.

From the phase evolution observed in *in situ* XRD measurements (**Fig. 3a, 3c**), we suspect two reactions that led to the nucleation of LTOPO (**Fig. 3e**):

$$\text{Li}_2\text{O} + 2\,\text{TiO}_2 + \text{P}_2\text{O}_5 \rightarrow 2\,\text{LiTiOPO}_4 \quad (500\,°\text{C}) \qquad \text{(R1)}$$

$$\text{LiTi}_2(\text{PO}_4)_3 + \text{Li}_3\text{PO}_4 + 2\,\text{TiO}_2 \rightarrow 4\,\text{LiTiOPO}_4 \quad (600\,°\text{C}) \qquad \text{(R2)}$$

R1 is associated with a large driving force of $\Delta G_{\text{rxn},o} = -279.27$ meV/atom at 500 °C, whereas R2 has a much weaker driving force of only $-40.22$ meV/atom at 600 °C

For R1, the ratio of nucleation rates between the two polymorphs is calculated to be $\frac{Q_o}{Q_t}(\text{R1}, 500°\text{C}) = 1.03 \times 10^{-17}$, suggesting that *t*-LTOPO nucleates much faster than *o*-LTOPO when starting from Li$_2$O, TiO$_2$, and P$_2$O$_5$. In contrast, for R2 the ratio is calculated to be $\frac{Q_o}{Q_t}(\text{R2}, 600°\text{C}) = 3.5 \times 10^{59}$, which instead suggests that *o*-LTOPO nucleates significantly faster than *t*-LTOPO when it forms from LiTi$_2$(PO$_4$)$_3$, Li$_3$PO$_4$, and TiO$_2$. These predictions agree well with the experimental observations. Since all other terms in Eqn. S3 remain unchanged between R1 and R2, reaction energy plays a determining role in the selectivity LTOPO polymorph.

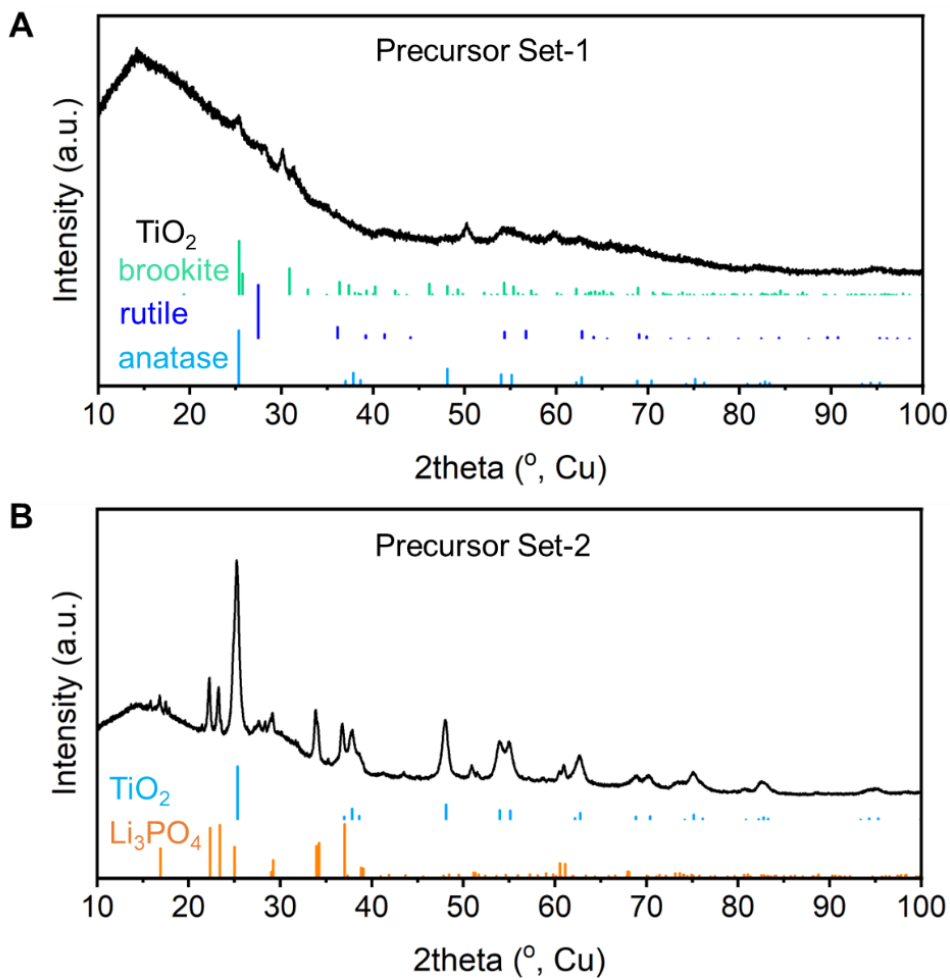

**Figure S1. XRD of ball-milled mixtures.** (A) After ball-milling, precursor *Set 1* ($Li_2CO_3$ + $TiO_2$ + $P_2O_5$) forms to amorphous solids and crystalline phase of $TiO_2$ polymorphs of brookite (ICSD#36408), rutile (ICSD#33837), and anatase (ICSD#121632). (B) After ball-milling, precursor *Set 2* ($Li_2CO_3$ + $TiO_2$ + $NH_4H_2PO_4$) transforms to crystalline $TiO_2$ (ICSD#121632) anatase and $Li_3PO_4$ (ICSD#10257).

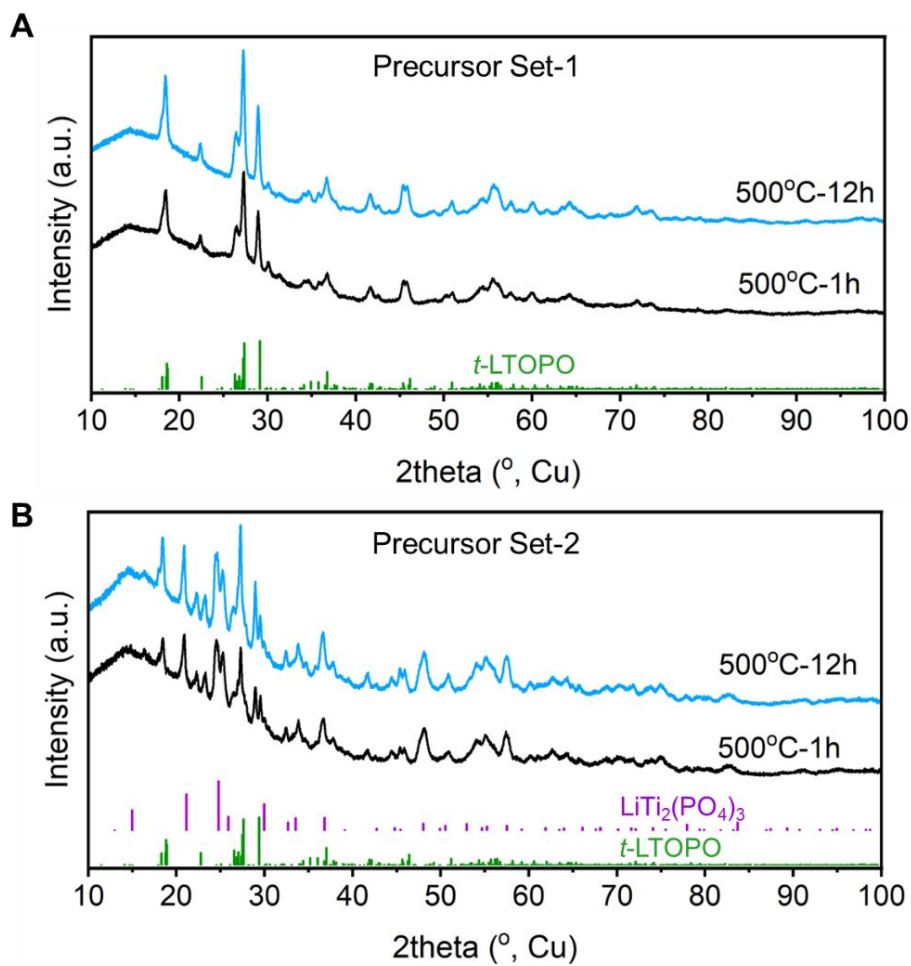

**Figure S2.** *Ex situ* XRD patterns and phase identification. (A) *t*-LTOPO (ICSD#39761) was formed as the single-phase by heating the precursor *Set 1* ($Li_2CO_3$ + $TiO_2$ + $P_2O_5$) at 500°C for 1 hour or 12 hours. (B) Mixed-phases containing *t*-LTOPO and rhombohedral $LiTi_2(PO_4)_3$ (ICSD#7930) were formed from heating *Set 2* ($Li_2CO_3$ + $TiO_2$ + $NH_4H_2PO_4$) at 500°C for 1 hour or 12 hours.

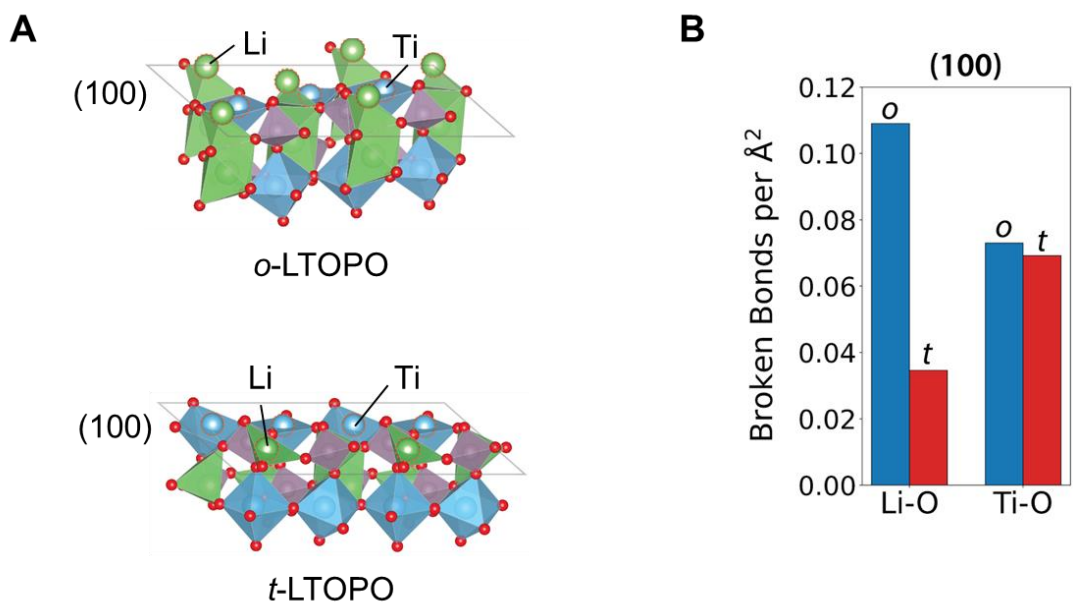

**Figure S3.** (A) Structure of the (100) surface for each polymorph, where the broken Li-O and Ti-O bonds are highlighted. (B) Populations of broken Li-O and Ti-O bonds on the (100) surfaces.

**References:**

1.  Mullin, J. W. in *Crystallization (Fourth Edition)* (ed J. W. Mullin) 181-215 (Butterworth-Heinemann, 2001).
2.  Langer, J. S. Metastable states. *Physica* **73**, 61-72, doi:https://doi.org/10.1016/0031-8914(74)90226-2 (1974).
3.  Sun, W. & Ceder, G. Induction time of a polymorphic transformation. *CrystEngComm* **19**, 4576-4585, doi:10.1039/C7CE00766C (2017).
4.  Kresse, G. & Furthmüller, J. Efficient iterative schemes for ab initio total-energy calculations using a plane-wave basis set. *Physical Review B* **54**, 11169-11186, doi:10.1103/PhysRevB.54.11169 (1996).
5.  Kresse, G. & Furthmüller, J. Efficiency of ab-initio total energy calculations for metals and semiconductors using a plane-wave basis set. *Computational Materials Science* **6**, 15-50, doi:https://doi.org/10.1016/0927-0256(96)00008-0 (1996).